\documentclass{icrc29}
\usepackage{graphicx,amssymb,amsmath,times}
\begin{document}
\title[]{Radio emission of highly inclined cosmic ray air showers measured with LOPES}
\author[J.Petrovic et al.] {J.~Petrovic$^g$, 
W.D.~Apel$^{a}$, 
F.~Badea$^{a}$,
L.~B\"ahren$^{b}$,
K.~Bekk$^{a}$, 
A.~Bercuci$^{c}$,
M.~Bertaina$^{d}$, 
\newauthor
P.L.~Biermann$^{e}$,
J.~Bl\"umer$^{a,f}$,
H.~Bozdog$^{a}$,
I.M.~Brancus$^{c}$,
S.~Buitink$^{g}$,
M.~Br\"uggemann$^{h}$,
\newauthor
P.~Buchholz$^{h}$,
H.~Butcher$^{b}$,
A.~Chiavassa$^{d}$,
K.~Daumiller$^{a}$, 
A.G.~de~Bruyn$^{b}$,
C.M.~de~Vos$^{b}$,
\newauthor
F.~Di~Pierro$^{d}$,
P.~Doll$^{a}$, 
R.~Engel$^{a}$,
H.~Falcke$^{b,e,g}$,
H.~Gemmeke$^{i}$, 
P.L.~Ghia$^{j}$,
R.~Glasstetter$^{k}$, 
\newauthor
C.~Grupen$^{h}$,
A.~Haungs$^{a}$, 
D.~Heck$^{a}$, 
J.R.~H\"orandel$^{f}$, 
A.~Horneffer$^{e,g}$,
T.~Huege$^{a,e}$,
K.-H.~Kampert$^{k}$,
\newauthor
G.W.~Kant$^{b}$, 
U.~Klein$^{l}$,
Y.~Kolotaev$^{h}$,
Y.~Koopman$^{b}$, 
O.~Kr\"omer$^{i}$, 
J.~Kuijpers$^{g}$,
S.~Lafebre$^{g}$,
\newauthor
G.~Maier$^{a}$,
H.J.~Mathes$^{a}$, 
H.J.~Mayer$^{a}$, 
J.~Milke$^{a}$, 
B.~Mitrica$^{c}$,
C.~Morello$^{j}$,
G.~Navarra$^{d}$,
\newauthor
S.~Nehls$^{a}$,
A.~Nigl$^{g}$,
R.~Obenland$^{a}$,
J.~Oehlschl\"ager$^{a}$, 
S.~Ostapchenko$^{a}$, 
S.~Over$^{h}$,
H.J.~Pepping$^{b}$,
\newauthor
M.~Petcu$^{c}$, 
T.~Pierog$^{a}$, 
S.~Plewnia$^{a}$,
H.~Rebel$^{a}$, 
A.~Risse$^{m}$, 
M.~Roth$^{f}$, 
H.~Schieler$^{a}$, 
\newauthor
G.~Schoonderbeek$^{b}$
O.~Sima$^{c}$, 
M.~St\"umpert$^{f}$, 
G.~Toma$^{c}$, 
G.C.~Trinchero$^{j}$,
H.~Ulrich$^{a}$,
\newauthor
S.~Valchierotti$^{d}$,
J.~van~Buren$^{a}$,
W.~van~Capellen$^{b}$,
W.~Walkowiak$^{h}$,
A.~Weindl$^{a}$,
S.~Wijnholds$^{b}$,
\newauthor
J.~Wochele$^{a}$, 
J.~Zabierowski$^{m}$,
J.A.~Zensus$^{e}$,
D.~Zimmermann$^{h}$\\
(a) Institut\ f\"ur Kernphysik, Forschungszentrum Karlsruhe,
76021~Karlsruhe, Germany\\
(b) ASTRON, 7990 AA Dwingeloo, The Netherlands \\
(c) National Institute of Physics and Nuclear Engineering,
7690~Bucharest, Romania\\
(d) Dipartimento di Fisica Generale dell'Universit{\`a},
10125 Torino, Italy\\
(e) Max-Planck-Institut f\"ur Radioastronomie,
53010 Bonn, Germany \\
(f) Institut f\"ur Experimentelle Kernphysik,
Universit\"at Karlsruhe, 76021 Karlsruhe, Germany\\
(g) Department of Astrophysics, Radboud University Nijmegen, 6525
ED Nijmegen, The Netherlands \\
(h) Fachbereich Physik, Universit\"at Siegen, 57068 Siegen, 
Germany \\
(i) Inst. Prozessdatenverarbeitung und Elektronik, 
Forschungszentrum Karlsruhe, 76021~Karlsruhe, Germany \\
(j) Istituto di Fisica dello Spazio Interplanetario, INAF, 
10133 Torino, Italy \\
(k) Fachbereich Physik, Universit\"at Wuppertal, 42097
Wuppertal, Germany \\
(l) Radioastronomisches Institut der Universit\"at Bonn, 
53121 Bonn, Germany \\
(m) Soltan Institute for Nuclear Studies, 90950~Lodz, 
Poland\\
      }

\presenter{Presenter: S. Buitink (sbuitink@astro.ru.nl)}

\maketitle

\begin{abstract}

LOPES (LOFAR Prototype Station) is an array of dipole antennas used for 
detection of radio emission from air showers.
It is co-located and triggered by the KASCADE (Karlsruhe Shower Core and 
Array Detector) experiment, which also provides informations about 
air shower properties. 
Even though neither LOPES nor KASCADE are completely 
optimized for the detection of highly inclined events,
a significant number of showers with zenith angle larger
than 50$^o$ have been detected in the radio domain, and many
with very high field strengths. Investigation of inclined showers can 
give deeper insight into the nature of primary particles that
initiate showers and also into the possibility that some of detected 
showers are triggered by neutrinos.  
In this paper, we show the example of such an event and present 
some of the characteristics of highly
inclined showers detected by LOPES.
\end{abstract}

\section{Introduction}

When a cosmic ray interacts with particles in the Earth atmosphere, it produces
a shower of elementary particles propagating towards the grounds with almost
the speed of light. The first suggestion that these air showers can produce radio
emission was given by Askaryan \cite{askaryan} 
based on a charge-excess mechanism. Recently, Falcke\&Gorham \cite{falcke-gorham}
proposed that the mechanism for radio emission of air showers is coherent geosynchrotron
radiation. Secondary electrons and positrons produced in the particle cascade are deflected
in the Earth magnetic field and this produces dipole radiation that is relativistically
beamed in the forward direction.
The shower front emitting the radiation has a thickness which is comparable to a wavelength
for radio emission below 100MHz (around few meters). The emission is coherent which amplifies the 
signal. 

Radio emission of cosmic ray air showers has been detected by LOPES (LOFAR Prototype Station) 
\cite{falcke-nature}, a phased array of dipole antennas co-located with
the KASCADE (Karlsruhe Shower Core and Array Detector) experiment which provides 
coincidence triggers for LOPES and well-calibrated informations about air-shower properties, like
electron number $N_e$, reconstructed muon number $N_{\mu}$, azimuth and zenith angle of the event.
The LOPES experiment and data reduction are described in detail by Horneffer et al. \cite{horneffer}.

Highly inclined showers are expected to be very well detectable in the radio domain
\cite{huege-falcke},\cite{gousset}. 
However, we have to mention that neither LOPES nor KASCADE are optimized for large
zenith angles. For example, KASCADE reconstruction of electron and muon number can be
not accurate especially in cases when the shower core of specific event falls out
of the KASCADE array. 

Inclined cosmic ray air showers are specific, since they travel trough few times longer
distances in the Earth atmosphere compared to vertical showers, and due to this most of the 
electromagnetic (particle) component of those showers has been 
absorbed. So, inclined 
showers that start high in the atmosphere (initiated by protons, iron nuclei or gamma-photons) will have
large electron deficiency on the ground level compared to vertical showers. 
On the other hand, neutrino induced showers may be generated at any distance from the 
ground \cite{capelle} so they could clearly
be distinguished from those whose primary particle initiated the shower
high in the atmosphere by the number of electrons that reach the ground level.


\begin{figure}[h]
\begin{center}
\includegraphics*[width=0.3\textwidth,angle=270,clip]{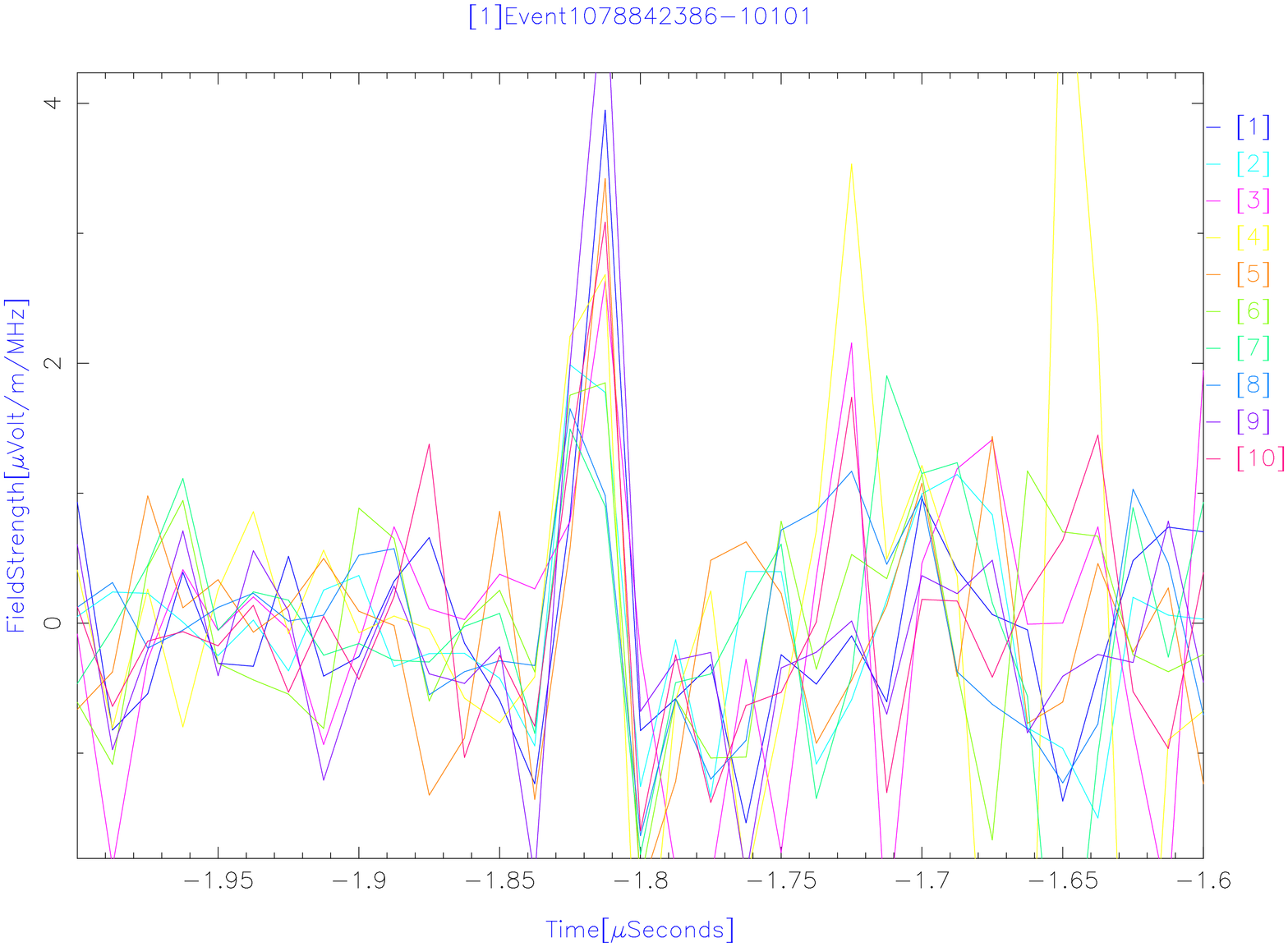}
\includegraphics*[width=0.3\textwidth,angle=270,clip]{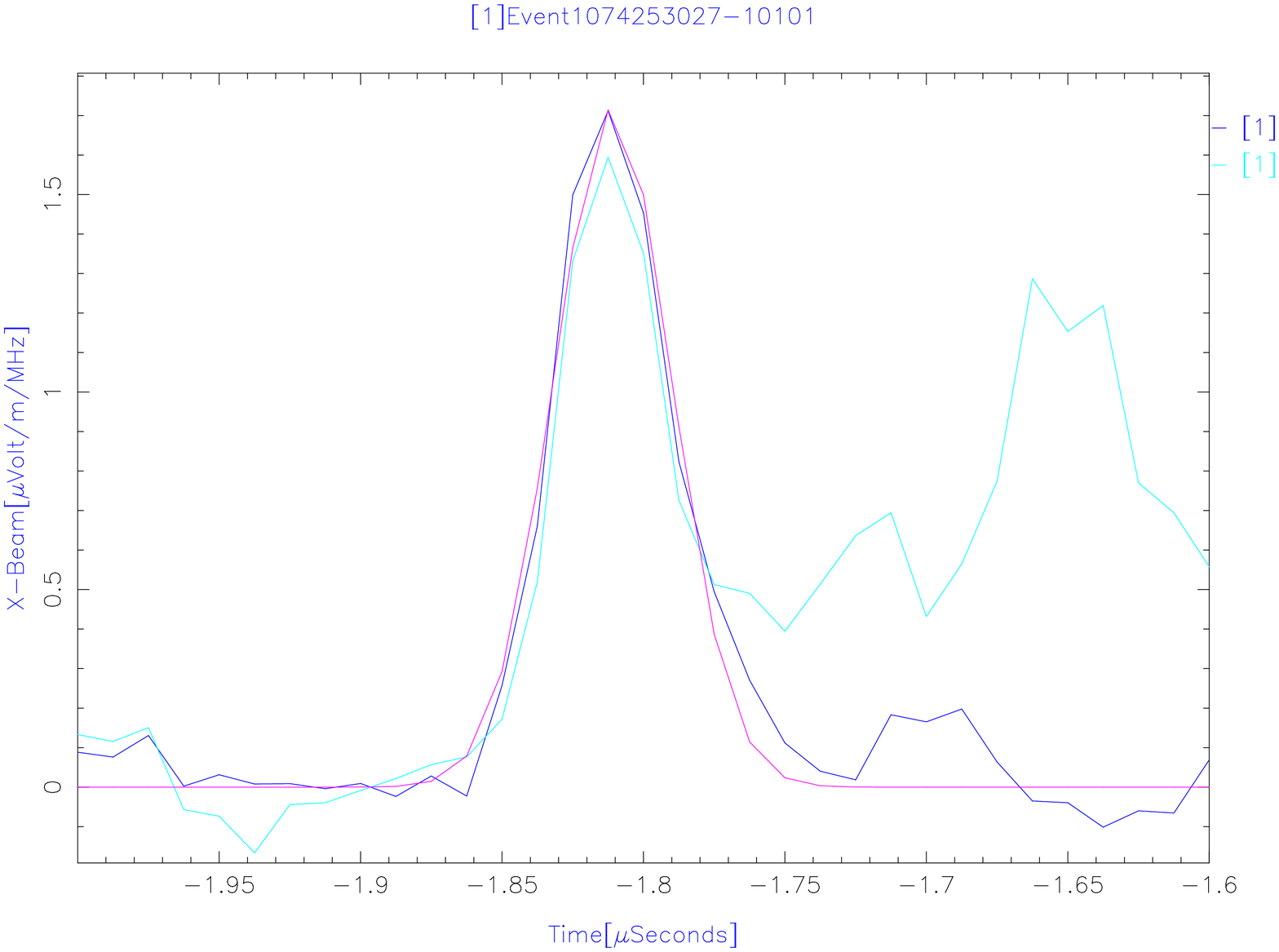}
\caption{\label{fig1}Left: Electric field as a function of time for 10 LOPES antennas for an event 
detected in March 2004 with zenith angle 53.3$^o$ and azimuth 
54.7$^o$. The angle between the shower core and the Earth magnetic field is 69.8$^o$.
$N_e$=1.5$\cdot$10$^6$, $N_{\mu}$=1.5$\cdot$10$^6$, reconstructed by KASCADE. Right:
Radio emission as a function of time after beam forming for the same event.
Dark blue line represents X-beam, light blue line the total power and 
pink line the Gaussian fit for X-beam.}
\end{center}
\end{figure}

\section{Discussion}

\begin{figure}[h]
\begin{center}
\includegraphics*[width=0.49\textwidth,clip]{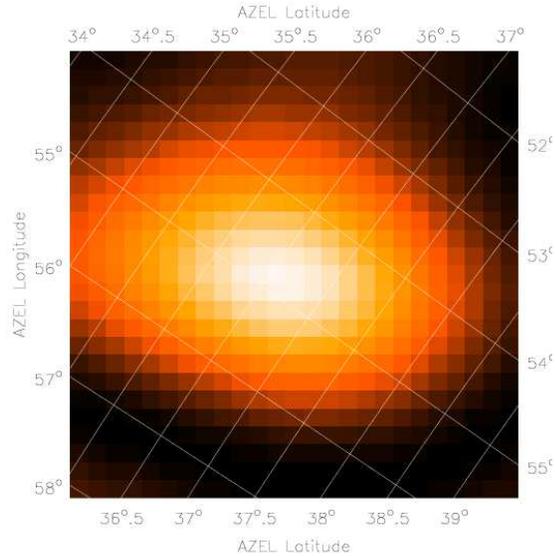}
\caption{\label{fig3}Radio map of the cosmic ray shower with zenith angle 53.3$^o$ detected in March 2004. 
Azimuth (AZEL longitude) and elevation (AZEL latitude) of the event are given on the axes.}
\end{center}
\end{figure}

We made a selection of inclined events from data taken during year 2004 (also detected by the KASCADE
array where the reconstructed shower cores were less than 100m away from the array center).
We found 2017 events with zenith angle larger than 50$^o$.
Then we introduced an additional condition: $N_{\mu}$$>$10$^5$.
In this way we narrowed the selection to 51 event and more than 40\% of those are detected
in the radio domain, many with very large field strengths, even though a threshold
on muon number is lower than the one used for bright events in \cite{falcke-nature}.
However, as we already mentioned, KASCADE is not optimized for large zenith angles, 
so the reconstruction of the electron and muon number failed for half of the detected events.
This leaves us with 10 events with strong radio signal and reliable shower properties
reconstructed by KASCADE.

As an example, we show here one of those events, detected in March 2004 with zenith angle 
53.3$^o$ and azimuth 54.7$^o$ with roughly the same number of electrons and 
muons $N_e$,$N_{\mu}$$\approx$1.5$\cdot$10$^6$ 
(reconstructed by KASCADE). The angle between the shower axis and the Earth magnetic field 
(geomagnetic angle) is 69.8$^o$.
 
In Figure 1 (left) we show the electric field as a function of time for each antenna. The field is coherent 
at -1.825$\mu$s which is the arrival time of the shower. The incoherent noise after is radio emission from
photomultipliers and in this case is very weak.
Figure 1 (right) shows the radio emission as a function of time after X(excess)-beam forming.
This beam is formed in the following way. First squared signals of all antennas are summed which gives
the total power. Then signals of all two antenna combinations
are multiplied and summed which gives the CC(cross correlation)-beam. Finally, CC-beam is multiplied 
with the ratio 
between CC-beam and total power. In this way a suppression of incoherent noise is achieved.

\begin{figure}[h]
\begin{center}
\includegraphics*[width=0.35\textwidth,clip]{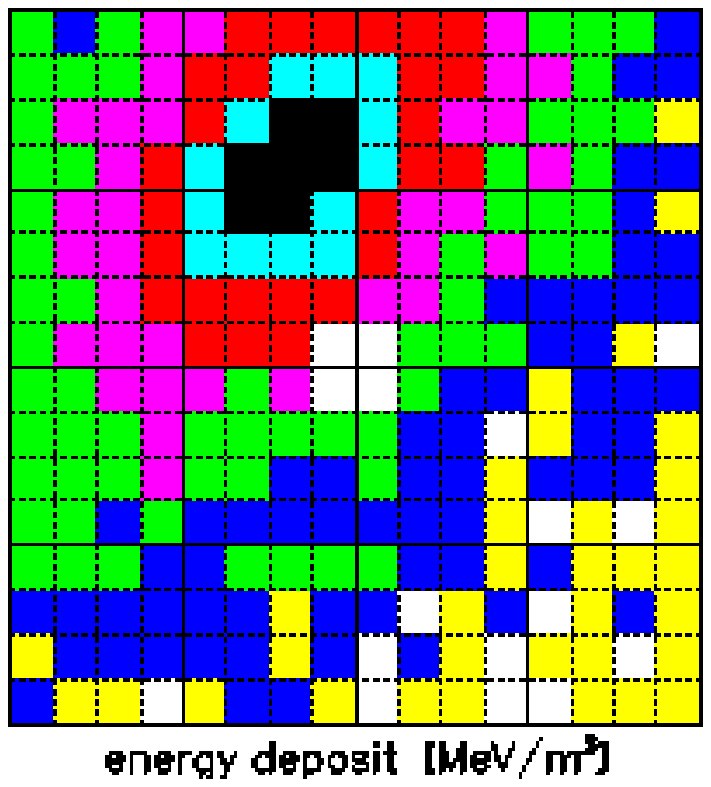}
\includegraphics*[width=0.35\textwidth,clip]{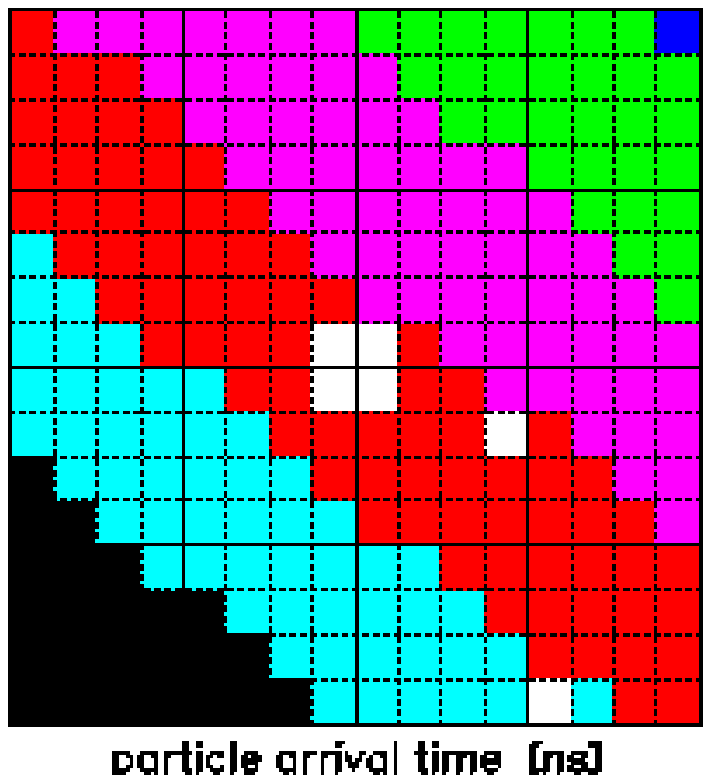}
\caption{\label{fig3}
Left: Energy deposit of the  
cosmic ray shower with zenith angle 53.5$^o$ detected in March 2004 for e/$\gamma$ KASCADE detectors.
Dark blue color shows energy deposit of 100Mev/m$^2$, red color 1000Mev/m$^2$. Maximum energy deposit
for this event is $\sim$4500Mev/m$^2$.
Right: Particle arrival time for the same event. Dark blue color represents $\sim$300ns, red $\sim$800ns.
}
\end{center}
\end{figure}

Figure 2 is a radio map of the example event. The air shower is the brightest point in the sky for
several tens of nanoseconds. The resolution of the map is $\sim$2$^o$ in azimuth and elevation towards the
zenith. 

Figure 3 gives the energy deposit of the chosen cosmic ray shower over the KASCADE array with 
e/$\gamma$ detectors
(left panel) and particle arrival time (right panel). 
We can see that the shower core falls within the KASCADE array and that
the maximum energy is deposited in the north-western part, within the LOPES array. We can 
notice elliptical shapes of isolines of energy deposit, which is typical for inclined events.

\section{Conclusions}

Even though neither LOPES nor KASCADE are completely 
optimized for the detection of highly inclined events,
we find that in a selection of events with zenith angle larger than
50$^o$ and $N_{\mu}$$>$10$^5$ (51 event)
around 40\% of all events is detected in the radio domain, and some of them 
with very high field strengths, like the example we have presented in this paper.
The most inclined cosmic ray air shower that we detected with LOPES has
a zenith angle of almost 70$^o$.

After checking the resonctruction of $N_e$ and $N_{\mu}$ different correlations 
can be considered, for example between radio pulse height and muon number, electron number or 
geomagnetic angle. This will also give more insight into the nature of primary particles that
initiate showers and the possibility that some of detected showers might have been triggered
by neutrinos.


\end{document}